\documentclass[english,aps,floats,showpacs,twocolumn,nofootinbib]{revtex4-1}

\usepackage[T1]{fontenc}
\usepackage[latin1]{inputenc}
\usepackage{graphicx}
\usepackage{epsfig}
\usepackage{psfrag}
\usepackage{verbatim}
\usepackage{enumerate}
\usepackage{rotating}
\usepackage{calc}
\usepackage{ifthen} 
\usepackage{color}

\definecolor{color8}{rgb}{0.5804, 0.0000, 0.8275}
\definecolor{color13}{rgb}{0.4039, 0.0275, 0.2824}
\definecolor{color18}{rgb}{0.2588, 0.2588, 0.2588}

\definecolor{rossos}{rgb}{0.8,0.2,0.3}

{\newcommand{\lsim}{\mbox{\raisebox{-.6ex}{~$\stackrel{<}{\sim}$~}}}
{ 
{
\newcommand{\be}{\begin{equation}}
\newcommand{\ee}{\end{equation}}
\newcommand{\bea}{\begin{eqnarray}}
\newcommand{\eea}{\end{eqnarray}}

\def\eV{{\rm \ eV}}
\def\GeV{{\rm \ GeV}}
\def\MeV{{\rm \ MeV}}

\def\eta{Y_B}
\def\XX{\mathcal X_i}

\def\timess{\hspace{-.5mm}\times\hspace{-.5mm}}

\begin{document}
\hspace{8cm}\rightline{ULB-TH/10-18}

\title{Re-reheating, late entropy injection and constraints from baryogenesis scenarios}
\author{Germano Nardini}
\author{Narendra Sahu}
\affiliation{Service de Physique Th\'eorique, Universit\'e Libre de Bruxelles, 1050 Brussels, 
Belgium}

\begin{abstract}
  Many theories of particle physics beyond the Standard Model predict
  long-lived fields that may have dominated the Universe at early
  times and then decayed. Their decay, which injects entropy in the
  thermal bath, is responsible for a second reheating, dubbed
  re-reheating, that could substantially dilute the matter-antimatter
  asymmetry created before. In this paper we analyze such late
  re-reheating and entropy dilution. It turns out that in some cases
  the usual analytic calculation badly fails if it is not rectified by
  some corrective factors that we provide.  We also determine the
  parameter space where the entropy dilution compromises models of
  baryogenesis. This region can be obtained by imposing some generic
  constraints that are applicable to any baryogenesis mechanism and
  long-lived field satisfying a few assumptions. For instance, by
  applying them to MSSM electroweak baryogenesis, thermal non-resonant
  leptogenesis and thermal resonant leptogenesis, we obtain that the
  initial abundances of long-lived fields with lifetime longer than
  respectively $5\times10^{13}, 10^{-2}$ and $10^{15}\,\GeV^{-1}$ are
  strongly constrained. Similarly, the same baryogenesis scenarios are
  incompatible with large oscillations of moduli with mass smaller
  than $\mathcal{O}(10^8), \mathcal{O}(10^{13})$ and $\mathcal
  O(10^{7})\GeV$ that are naturally coupled to the visible sector via
  gravitational dimension-five operators.

\end{abstract}
\pacs{98.80.Cq}
\maketitle

\section{Introduction}
%
The history of the Universe before Big Bang Nucleosynthesis (BBN) is
an open issue. The standard cosmological scenario considers epochs
prior to BBN to be radiation dominated. However, this assumption is
questionable since experiments cannot put severe constraints on the
Universe at temperature $T\gg \mathcal{O}(1\,$MeV) and moreover
several theoretical frameworks predict a modification of the standard
cosmological picture. For instance, models having long lived fields
that are not in thermal equilibrium may induce an era where the
radiation energy is subdominant.  Potentially, this epoch might emerge
in presence of flat directions~\cite{flat_decay},
Q-balls~\cite{qball_decay}, gravitinos~\cite{gravitino_decay},
axinos~\cite{axino_decay}, moduli~\cite{moduli_decay}, which we will
generically refer to as $X$ field in the following.

On the other hand, the history of the Universe at $T\lesssim
\mathcal{O}(1\,$MeV) is well established: any primordial $X$-dominated
epoch must end before the onset of BBN in order not to jeopardize the
predictions of the primordial element
abundances~\cite{Hannestad:2004px, Jedamzik:2009uy}.  The required
return to the radiation dominated era may occur via $X$ decay. This
process, which we dub {\it re-reheating} to distinguish it from the
first reheating happened in the inflationary epoch, dumps entropy into
the thermal bath and may considerably dilute the pre-existing species.

Various experimental observations can be explained by
re-reheating. Among many we mention the possibility of tuning the
$X$-decay rate to generate the right amount of non-thermal cold dark
matter required for structure
formation~\cite{nonthermal_DM}. Moreover, the $X$ decay can be invoked
to wash out unwanted relics of early stages of the Universe
({\it{e.g.}}, monopoles and domain walls), to circumvent the gravitino
problem~\cite{Hasenkamp:2010if} or to reduce an overabundance of
thermal dark matter candidates~\cite{yaguna}.

All these possibilities highlight that future detection of new
particles might provide a link between (non-standard) cosmology and
particle physics that might be misunderstood if non classical
cosmological scenarios are not taken into account. Thus, in order to
infer cosmological issues from particle experiments, we will need to
clarify as much as possible the picture of the Universe before BBN. In
particular, bounds on the possible $X$-dominated epoch will be
useful~\footnote{Some constraints on $X$ are deduced by noticing that
  the $X$ decay might have left a trace in the Cosmic Microwave
  Background (CMB)~\cite{curvature_perturbation}. Unluckily, such a
  signature cannot be disentangled from the inflaton one so that the
  derived bounds on $X$ depend strongly on the assumed inflationary
  model.}.  With this goal in mind, in the present paper we link the
puzzle of the observed Baryon ($B$) Asymmetry of the Universe (BAU)
with the $X$ decay. In this manner we obtain some bounds on models
involving baryogenesis and late-time entropy injection where the
latter does not induce $B$ violations.

The experimental measure of the BAU is achieved by the BBN~\cite{pdg}
and CMB~\cite{WMAP} analyses. Under the assumption that only photons
and neutrinos are relativistic at $T\!\lesssim\!{\mathcal O}(1\!\MeV)$,
these analyses imply
\bea
\label{BBN} 
7.2<&\eta^{exp} \times 10^{11}&<9.2 \quad {\rm ~~BBN~ at~ 95\%~
  C.L.}~,\\
\label{WMAP}
8.4<&\eta^{exp} \times 10^{11}&<9.2 \quad {\rm ~~CMB~ at~ 95\%
  ~C.L.}~, 
\eea
where $\eta^{exp}\equiv N_B^*/S^* $ and $S^*$ ($N_B^*$) is the total
entropy (total baryon minus antibaryon number) during the BBN epoch,
{\it i.e.}, $T\lsim T_{BBN}\equiv 4$ MeV
\cite{Hannestad:2004px}. These measures constrain the late-time
evolution of $\eta(t)\equiv N_B(t)/S(t)$ and, once $N_B(t)$ is known,
some bounds on $S(t)$ can be inferred. In other words, by knowing when
and in what abundance the $B$ asymmetry was produced ({\it i.e.}, by
assuming a given baryogenesis mechanism), one can constrain the
$X$-decay entropy injection that is compatible with the above BBN and
WMAP measures. Alternatively, by assuming some characteristics of the
$X$ field, one can impose extra requirements on the mechanism
responsible for the BAU.

Since we perform our analysis without considering any specific
baryogenesis mechanism or any particular $X$ field, our derived bounds
are easily applicable to a wide class of models. For this reason, the
generic results we obtain can be used as tools by which the reader can
easily estimate the parameter region where her/his favorite
baryogenesis mechanism and entropy injecting field (fulfilling some
requirements explained in the text) are compatible.

As illustrative applications, we show the implications of our results
for some specific baryogenesis mechanisms. More precisely, we obtain
the parameter region of a generic $X$ field where the late-time
entropy injection is compatible with electroweak baryogenesis and
resonant or non-resonant thermal leptogenesis embedded in the Minimal
Supersymmetric Standard Model (MSSM). Subsequently we consider the
specific case of $X$ fields being moduli.

The paper is organized as follows. In Section II we quantify
analytically the entropy injection produced by the $X$ decay and
compare it with the numerical prediction. Section III is devoted to
demonstrate how the different baryogenesis mechanisms constrain the
lifetime and energy density of a generic $X$ field and viceversa.  In
Section IV we apply our results to concrete cases of baryogenesis
mechanisms and long-lived $X$ fields.  Finally, we leave Section V for
conclusions and Appendix for technical details.

\section{Quantifying entropy injection}
%
%
%
Let us assume the existence of a weakly-coupled field $X$ with
lifetime $\tau$ and energy density $\rho_X$ redshifting as
$R(t)^{-3}$, where $R(t)$ is the expansion scale factor of the
Freedman-Robertson-Walker Universe. Before the $X$ decay, which we
assume to be mostly into components of the thermal bath, the radiation
energy density $\rho_R$ falls faster then $\rho_X$ by a factor $R(t)$
roughly. Hence, if the $X$ decay is suppressed enough, a $X$-dominated
epoch emerges even though initially $\rho_X$ is subdominant to
$\rho_R$. This deviation from the standard history of the Universe
ends once the $X$ fields decay away, what may release a sizable amount
of entropy into the thermal bath. In this section we quantify this
entropy injection and the temperature at which this injection ends.

\subsection{Setup and numerical evolution}
\label{sec:Num}
The evolution of the exponentially decaying $X$ fields with energy
density scaling as $R(t)^{-3}$ is described by the
equation~\footnote{It is assumed that the $X$ field has a number
  density much larger than its equilibrium value, or to be an unstable
  field, when $3 H \tau\lesssim 1$.}~\cite{kolb}
\be
\label{bltz} 
\dot\rho_X + 3 H \rho_X= - \rho_X / \tau ~. 
\ee
Its solution is
\be
\label{rhoX}
\rho_X(R)=\rho_X(R_i)\, \frac{R_i^3}{R^3}\, \exp(-t/\tau) ~,
\ee
where $R_i$ is the scale factor at some initial time $t_i\ll\tau$ and
$H$ is the Hubble constant given by
\be
\label{H}
H^2=(\dot R/R)^2= (\rho_X + \rho_R)/(3 m_P^2)~, \ee
with the reduced Planck mass $m_P=2.4\times 10^{18}\GeV$.  Assuming
all the $X$ decay products to thermalize sufficiently fast implies
that the total entropy $S$ evolves as
\be
\label{S}
\dot S = \left( \frac{2 \pi^2 g(t)}{45~ S}\right)^{1/3} R^4 ~\rho_X/\tau ~,
\ee
where $g(t)$ is the number of relativistic degrees of freedom in the
thermal bath. In turn, $S$ is linked to $\rho_R$ by the expression
\be
\label{rhoR}
\rho_R=\frac{3}{4} \left(\frac{45~S^4}{2\pi^2 g(t)}\right)^{1/3} R^{-4} ~.
\ee
Therefore, the entropy dilution caused by the $X$ decay,
\be
\label{DeltaDef} 
\Delta\equiv \frac{S(t\gg\tau)}{S(t_i)}~, 
\ee
can be obtained by solving numerically the coupled
Eqs.~(\ref{rhoX})-(\ref{rhoR}) and imposing the initial conditions
$\rho_X(R_i) = \rho_X^i$ and $\rho_R(R_i)=\rho_R^i$.

In the numerical solution we implement $g(T)$ assuming that the QCD
phase transition takes place at $T\approx 0.2$\,GeV
\cite{kolb}. Moreover, the time dependence in $g(t)$ required in
Eq.~(\ref{S}) is derived iteratively from $g(T)$ by solving the above
differential equations and determining $T(t)$ from $S(t)=2\pi^2 g(T)
T^3 R^3/45$.

As an example, in Fig.~1 (dashed lines) the numerical evolution of
$\rho_R(t)$, $\rho_X(t)$ and $S(t)$ in the Standard Model (SM)
are shown for initial conditions $\rho_R^i/\rho_X^i=10^{10},
T_i=10^{9}$\,GeV and $\tau=10^{22}$ GeV$^{-1}$. The corresponding
dilution factor turns out to be $\Delta\simeq 7.8$.

\begin{figure}
$~t_e$ \hspace{3.35cm} $\,t_d$ \vspace{-.95cm}
\begin{center}
\includegraphics[width=0.49\textwidth]{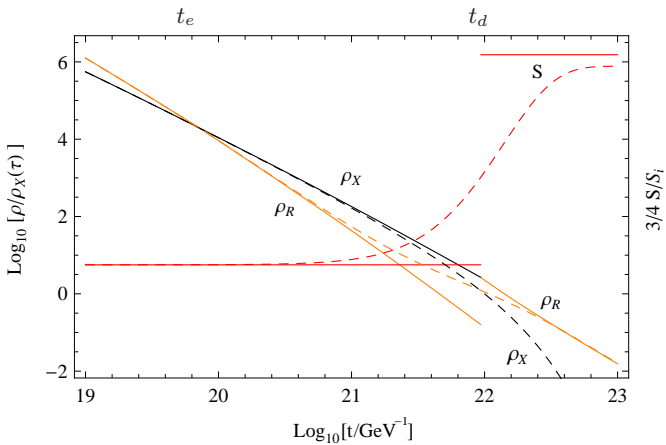}
\caption{Evolution of the energy densities $\rho_R$ and $\rho_X$ and
  the total entropy $S$ as function of time for initial conditions
  $\rho_R^i/\rho_X^i=10^{10}, T_i=10^{9}$\,GeV and $\tau=10^{22} $
  GeV$^{-1}$. The result of the numerical (analytical) solution is
  shown by dashed (solid) lines. The $X$-dominated epoch starts at
  $t_e$ and finishes at $t_d$.}
\label{figure-1}
\end{center}
\end{figure}

\subsection{Analytic approximation}
\label{AnApp}
A straightforward analytic approximation to estimate the entropy
dilution $\Delta$ is based on simplifying the previous exponential
decay to an instantaneous one occurring at the time $t_d\simeq\tau$
\cite{kolb}~\footnote{The present fully-analytic approach is the most
  straightforward for practical purposes, so we will compare our
  numerical results with it. For some semi-analytic approximations
  see Ref.~\cite{Scherrer:1984fd}.}. This approximation implies
$S(t)=S_i$ for $t<t_d$ and it yields an evolution of the Universe as
the one described by solid lines in Fig.~1, as we prove now.

Let us assume the Universe to be radiation dominated at the initial
time $t_i$~\footnote{In practice other reliable situations can be led
  to this case by shifting and redefining the time.}. For large enough
$\tau$ there exists an equilibrium time $t_e$ ($\gg t_i$) after which
$\rho_R$ becomes smaller than $\rho_X$.  Using $S_i=S_e$ (quantities
with the index $i,e,d,rh$ are evaluated at $t=t_i,t_e,t_d^-,t_d^+$,
respectively) one gets
\be
\label{Te}
\rho_R^{e}=\rho_X^{e}= \frac{g_e T_e^3}{g_i T_i^3} \rho_X^i \quad
\Rightarrow \quad T_e=T_i \rho_X^i/\rho_R^i~.
\ee

By taking the total energy and the scale factor $R$ to be constant at
the moment of the decay, it turns out to be
\be
\label{en_conserv}
\rho^d_X\equiv \rho_X(t_d^-)=\rho^{rh}_R\equiv
\rho_R(t_d^+)~,
\ee
and subsequently
\be
\label{rhoRH}
\rho_R^{rh}=\frac{g_d T_d^3}{g_e T_e^3} \rho_R^e \quad \Rightarrow
\quad T_{rh}^4= \frac{g_d}{g_{rh}} T_d^3 T_e ~, 
\ee
which implies an (apparent) instantaneous growth of
$T$~\footnote{Alternatively, the instantaneous decay could be treated
  by maintaining fixed $T$ and increasing instantaneously $R$. This
  would not lead to different conclusions.}.

The time $t_d$, at which the sudden decay produces the discontinuity,
is related to $\tau$ by the definition
\be
\label{timedecay}
\frac{3}{2} H[\rho_X^d]\equiv \tau^{-1}~, 
\ee
where the chosen prefactor $3/2$ is motivated by the matter-like
dominated epoch prior to the decay. Correspondingly, at $t=t_d$ the
$X$ decay increases the entropy by the factor
\be
\label{Delta}
\Delta=\frac{S_{rh}}{S_d}=\frac{g_{rh} T_{rh}^3}{g_d T_d^3}=\frac{T_e}{T_{rh}}~,
\ee
and, since Eqs.~(\ref{en_conserv}) and (\ref{timedecay}) imply
\be
\label{Trh}
T_{rh}^4=\frac{40 ~m_P^2}{\pi^2 g_{rh} \tau^2}~, 
\ee
one concludes
\be
\label{DeltaAnaly}
\Delta=1.38\times10^{-9}~\frac{\rho_X^i}{T_i^3}~
\frac{g_{rh}^{1/4}}{g_i} ~\left(\frac{\tau}{\GeV^{-1}}\right)^{1/2}~. 
\ee

In the rest of the paper we will use Eqs.~(\ref{Trh}) and
(\ref{DeltaAnaly}) as analytic estimates of the re-reheating
temperature and entropy dilution. 

Notice that Eq.~(\ref{Delta}) and thus (\ref{DeltaAnaly}) are strongly
sensitive to the evaluation of $T_e$ which derives from the equality
$S_i\!=\!S_e$ that is valid only for $t_e\!\ll\!\tau$. Hence for
$\Delta\approx1$, {\it i.e.} $T_e\simeq T_{rh}$,
Eq.~(\ref{DeltaAnaly}) is unreliable. On the other hand,
Eq.~(\ref{DeltaAnaly}) turns out to be realistic for $\Delta\gg 1$, as
Fig.~1 suggests and as we will show in the next section.
 
Fig.~1 also highlights that the sudden decay approximation is not
useful to describe quantities at $t\lesssim\tau$ at which the
discrepancy with the numerical quantities is huge.  In particular, it
is well known that the discontinuity of $T$ at $t=t_d$ is an artifact
of the crude approximation: actually the temperature of the Universe
always decreases smoothly during the decay~\cite{Scherrer:1984fd}, as
the numerical analysis in Fig.~1 highlights.

Finally, notice that in order to compare the evolutions of the
analytical and numerical quantities in Fig.~1, the different scale
factors $R(t)$ relative to each one of the two approaches are
introduced.  This allows to observe that in the numerical evolution
$\rho_X$ returns subdominant near $t=\tau$, what is reproduced in the
analytic approximation thanks to the factor 3/2 chosen in the
definition~(\ref{timedecay}). If this factor were lowered, the instant
of the temperature discontinuity would be anticipated.

\subsection{Entropy production and re-reheating temperature}
\label{entropy-rereh}
Besides the entropy dilution, the estimates of the re-reheating
temperature $T_{rh}$ obtained by the analytic and numerical approaches
are important. However, while in the sudden decay approximation
$T_{rh}$ is well defined -- it is the temperature just after the decay
-- in the numerical method it is not. In order to extend the concept
of re-reheating temperature to the latter case we introduce the
convention considering that the entropy injection ends when its
remaining variation $\xi$ is tiny. In this way in the numerical
analysis we can define implicitly $T_{rh}$ as $S(T\ll
T_{rh})/S(T_{rh}) \equiv 1+\xi$ and, after an opportune choice of
$\xi$, we can use the bound
\be
\label{tooConser}
T_{rh}\geq T_{BBN}~,  
\ee 
to avoid alterations of BBN predictions.

To determine $\xi$ we observe that 1\% of entropy injection after
$T_{BBN}$ roughly corresponds to the resolution of the measurements
(\ref{BBN}) and (\ref{WMAP}). This means that even in the pathological
parameter scenario yielding $\eta(T=T_{BBN})\simeq 8.4\times 10^{-11}$
[the lower bound allowed by (\ref{BBN}) and (\ref{WMAP})], a further
1\% dilution does not lower appreciably $\eta$ which then remains
within the experimental constraint at any $T\le T_{BBN}$. For this
reason, the requirement of not jeopardizing the BBN predictions can be
safely replaced by the bound (\ref{tooConser}) with $\xi=0.01$ (see
Appendix for considerations about other choices of
$\xi$). Consequently, the definition of reheating temperature that we
will use in the numerical analysis will be
\begin{equation}
\label{Trh_def}
\frac{S(T\ll T_{rh})}{S(T_{rh})}\equiv 1.01~.
\end{equation}

\subsection{Analytical versus numerical\label{comparison}}
%
\begin{figure}
\begin{center}
\includegraphics[width=0.47\textwidth]{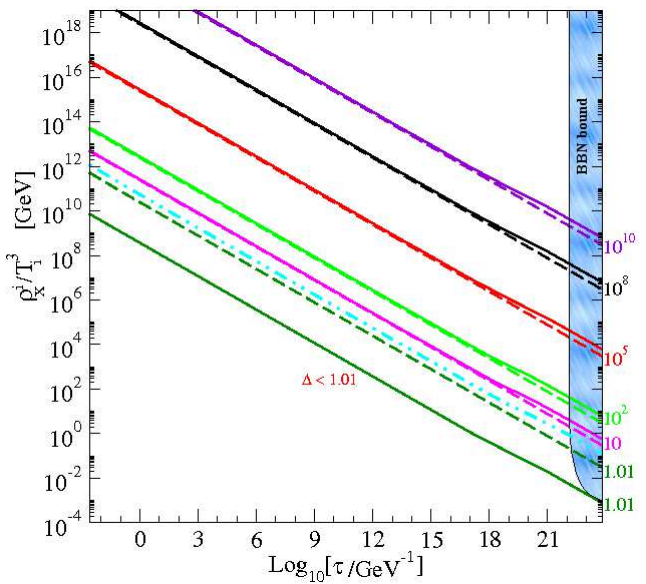}
\caption{Contours plot of $\Delta =1.01, 10, 10^2, 10^5, 10^8,
  10^{10}$ (labels on the right) as function of $\rho^i_X/T_i^3$ and
  $\tau$ determined analytically (dashed line) and numerically (solid
  line) using SM degrees of freedom. Below the dotted-dotted-dashed
  curve the $X$ decay injects entropy even without any $X$-dominated
  epoch. The filled region is excluded by BBN.}
\label{figure-2}
\end{center}
\end{figure}
In Fig.~\ref{figure-2} we show the numerical (solid) and analytical
(dashed) contour lines of $\Delta$ as function of $\tau$ and
$\rho_X^i/T_i^3$. The analytical lines are produced with $g(T)=106.75$
for simplicity while the numerical curves are calculated with the
exact $g(T)$ of the SM \cite{kolb}. As expected from
Eq.~(\ref{DeltaAnaly}), once $T_i$ is high enough to guarantee the
initial radiation-dominated epoch, $\Delta$ is basically independent
of it for $\rho_X^i/T_i^3$ and $g_i$ fixed.

We also determine numerically the parameter space violating the BBN
bound (\ref{tooConser}). This region corresponds to the filled area of
Fig.~2, whose border curve $T_{rh}=T_{BBN}$ reaches the line
$\Delta=1.01$ asymptotically.  This constraint on $\tau$ is by one
order of magnitude stronger than the one obtained analytically by
Eq.~(\ref{Trh}).

The numerical analysis highlights the existence of small entropy
dilution even in absence of a $X$-dominated epoch. Indeed below the
dotted-dotted-dashed line of Fig.~\ref{figure-2} the $X$ field decays
before $\rho_X$ reaches $\rho_R$ but produces anyway a dilution
$\Delta\lesssim 2.8$. Approaching this regime corresponds to
considering $t_e\simeq \tau$ for which the analytic approximation
fails, as we already have stated in Section~\ref{AnApp}. As a
consequence, the constraint $\Delta<1.01$ evaluated analytically
allows for values of $\rho_X^i/T_i^3$ that actually are excluded by
two order of magnitude.  Instead for
$\Delta=10,10^2,10^5,10^8,10^{10}$ the analytical and numerical curves
are in good agreement and the difference appearing at large $\tau$ is
due to the simplification $g(t)=106.75$ in the analytic estimates.

\begin{figure}
\begin{center}
\includegraphics[width=0.47\textwidth]{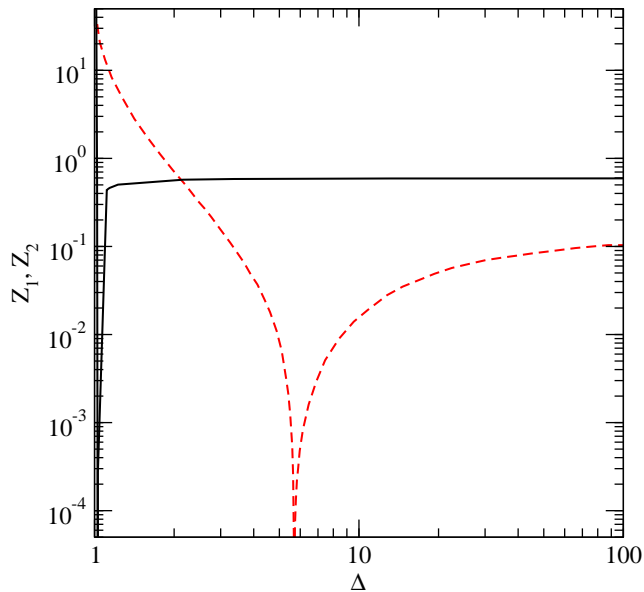}
\caption{The corrective factors $Z_1 \equiv
  |T_{rh}^{(num)}/T_{rh}^{(ana)}-1|$ (solid line) and $Z_2 \equiv
  |(\rho_X^i/T_i^3)^{(ana)}/(\rho_X^i/T_i^3)^{(num)}-1|$ (dashed line)
  for a fixed $\tau$ as a function of $\Delta $.  The argument in the
  modulus of $Z_1$ ($Z_2$) is negative for $\Delta\gtrsim1$ (positive
  for $\Delta\gtrsim6$).}
\label{figure-3}
\end{center}
\end{figure}

At this point one might be interested in knowing by what factors one
should correct the analytic results in order to reproduce the
numerical outcomes. These factors can be extracted from
Fig.~\ref{figure-3} where the absolute percentage errors
$$Z_1 \equiv |T_{rh}^{(num)}/T_{rh}^{(ana)}-1|~,$$\\[-1.1cm]
$$Z_2 \equiv|(\rho_X^i/T_i^3)^{(ana)}/ (\rho_X^i/T_i^3)^{(num)}-1|~, $$
are plotted as function of $\Delta$ (at $\tau = 10^{14}\,{\rm
  GeV}^{-1}$ but $Z_1$ and $Z_2$ are essentially independent of
$\tau$). It turns out that the analytic approach overestimates
$T_{rh}$ by a factor $\beta\simeq2.5$, typically. Moreover for a given
value of $\rho_X^i/T_i^3$, it underestimates the entropy dilution when
$\Delta \lesssim 6$ and its error remains below $\sim10\%$ when
$\Delta\gtrsim 3$. In particular, the error in estimating the dilution
by Eq.~(\ref{Delta}) or (\ref{DeltaAnaly}) cannot be attributed
univocally to a mismatch in $T_{rh}$ [which might be due to a wrong
choice of the prefactor in Eq.~(\ref{timedecay})] because in such a
case the equality $Z_1=Z_2$ should always arise ($T_e$ is evaluated
very precisely in both approaches for $\Delta\gg1$).

%


\section{Baryogenesis bounds}
\label{III}
The formation of primordial elements requires $\eta(t)$ at BBN times
to be compatible with the measures (\ref{BBN}) and (\ref{WMAP}). In
the previous section we used these bounds to constrain the history of
$S(t)$ by assuming $N_B(t)$ constant at $T\lesssim T_{BBN}$. However,
further constraints on the $X$ decay can be inferred by knowing the
evolution of $N_B(t)$ before BBN, as we explain now.

Let us consider a given baryogenesis mechanism that at the temperature
$T_B$ generates a $B$ asymmetry $\eta^{max}$ at best. Depending on the
relative times at which the $B$ asymmetry and entropy dilution are
produced, three different cases are possible:
\begin{enumerate}
\item the entropy injection occurs {\it exclusively after} $T_B$ and
  thus the condition $\Delta\leq \eta^{max}/\eta^{exp}$ is required;
\item the entropy injection happens {\it exclusively before} $T_B$
  ({\it i.e.}, $T_{rh}>T_B$) and then any $\Delta$ is allowed;
\item the period of entropy injection {\it encloses} $T_B$ and hence
  the part of entropy produced after $T_B$ must be smaller than
  $\eta^{max}/\eta^{exp}$~\footnote{To avoid confusion in the
    definitions, we stress that $\Delta$ is the dilution due to the
    {\it whole} entropy produced by the $X$ decay, while
    $\tilde\Delta$ is only that {\it portion} of dilution occurring
    {\it exclusively after} $T_B$.}:
\be
\label{tDelta}
\tilde\Delta(T_B)\equiv \frac{S(T \ll T_{rh})}{S(T_B)} \leq
\eta^{max}/\eta^{exp}~.  
\ee
\end{enumerate}

\begin{figure}
\begin{center}
\includegraphics[width=0.47\textwidth]{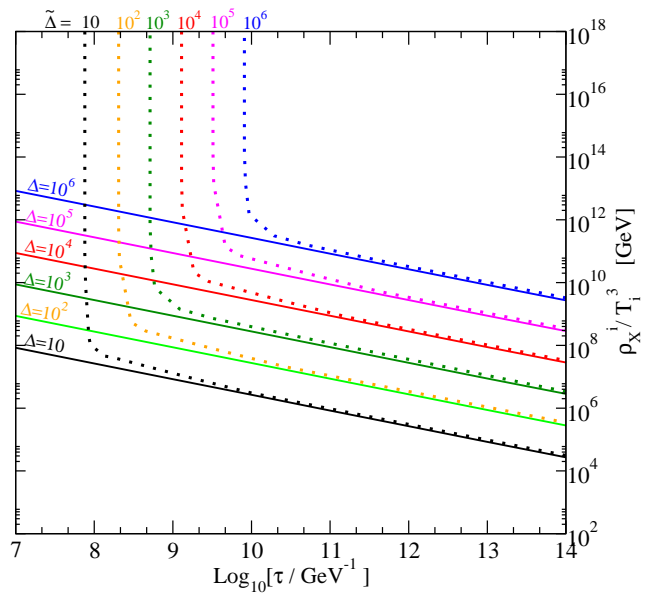}
\caption{Contour plots of the bounds $\Delta$ and
  $\tilde\Delta(T_B=10^5\GeV)$ for
  $\eta^{max}/\eta^{exp}\hspace{-3pt}=\hspace{-3pt}$
  $10$,$10^2$,$10^5$,$10^6$ are marked by solid and dotted lines,
  respectively. The region excluded by each bound stands above the
  corresponding curve.}
\label{figure-4}
\end{center}
\end{figure}

Notice that the parameter space fulfilling Eq.~(\ref{tDelta}) includes
the region contemplated by the possibilities 1 and 2. Therefore the
parameter region where a given baryogenesis mechanism is compatible
with the $X$ decay is determined by the bound (\ref{tDelta}).

As example, in Fig.~4 we present the $\tilde\Delta$ constraints
(dotted lines) for some baryogenesis mechanisms generating
$\eta^{max}/\eta^{exp}=10,10^2,10^5,10^6$ at $T_B=10^5$ (the figure
also reports the numerical lines $\Delta=10,10^2,10^5,10^6$ in
solid). The parameter space excluded by each $\tilde\Delta$ bound
stands on the right of the corresponding line~\footnote{In realistic
  cases $T_B$ stands for the temperature interval during which the $B$
  asymmetry is created. Here we work in the limit that the entropy
  produced during this interval is negligible. On the contrary, the
  vertical part of the $\tilde \Delta$ curves is stumped while its
  oblique section holds sharp.}.

The curves $\tilde \Delta$ and $\Delta$ overlap asymptotically because
at large $\tau$ the entropy is injected mostly after $T_B$ and hence
we have $S(T_B)\simeq S(T_i)$ in Eqs.~(\ref{DeltaDef}) and
(\ref{tDelta}). Thus, in this regime the parameter space excluded by
Eq.~(\ref{tDelta}) can be derived analytically by the condition
$\Delta\leq \eta^{max}/\eta^{exp}$.

Instead, in the region where the curves $\tilde\Delta$ are vertical
there is no bound on $\Delta$ but only on $\tau$. In such a region the
$\tilde\Delta$ condition is much less stringent than naively requiring
$T_{rh}\ge T_{B}$ (which actually corresponds to $\tilde\Delta\simeq
1$). The reason is that big values of $\eta^{max}/\eta^{exp}$ allow
for large dilutions of the produced $B$ asymmetry so that a region
with $T_{rh}< T_{B}$ is permitted.  This is quantified in Table I
where for various values of $\eta^{max}/\eta^{exp}$ we report the
bounds on $t_B/\tau$ and $T_{B}/T_{rh}$ valid in the regime of very
large $\Delta$,
that is, when the curve $\tilde\Delta$ is vertical.

The interesting issue is that the values in Table I are basically
independent of variation on $t_B$ and $T_B$. Moreover the shape of the
curves of the bounds $\tilde\Delta$ is universal, as we see in Fig.~4
where every dotted line could be obtained by the translation of one of
the others. This implies that by our results anyone can take some
long--lived field and baryogenesis mechanism and determine their
compatibility without implementing further numerical analyses. To do
it, one needs only to locate the curve (\ref{tDelta}) in the plan
$\tau$--$\rho_X^i/T_i^3\,$ for the desired framework. This can be
achieved as follows.

%
\begin{table}
\footnotesize
\begin{tabular}{ | c | c | c |c|c|c|c|}
    \hline
    $\tilde\Delta<$ &  $10$ & $10^{2}$ & $10^{3}$ & $10^{4}$
    & $10^{5}$ & $10^{6}$ \\ 
    \hline
    $t_B/\tau\gtrsim$ &  $2\timess10^{-1}$ & $3\timess10^{-2}$ & $5\timess10^{-3}$ & $7\timess10^{-4}$
    & $1\timess10^{-4}$ & $2\timess10^{-5}$ \\ 
    \hline
    $T_B/T_{rh}\hspace{-.5mm}\lesssim$ &  $3.4$ & $5.5$ & $8.8$ & $14$ & $21$ & $33$ \\
    \hline
    \end{tabular}
    \caption{Numerical evolution of the bound on $\tau$ ($T_{rh}$) as function of $\tilde
      \Delta$ and $t_B$ ($T_B$) in the regime of very large
      $\Delta$.}
\end{table}
\normalsize

First of all, one has to obtain the horizontal position of the
$\tilde\Delta$ curve. This is provided by the constraint on $\tau$ in
presence of very large $\Delta$ (namely, the position of the vertical
part of the $\tilde\Delta$ curve). It can be found by Eq.~(\ref{Trh})
relating $\tau$ to the analytic estimate of $T_{rh}$ which in turn
is constrained by
\be
\label{baryoB}
T_{rh}\gtrsim\frac{\beta}{\gamma}T_B\qquad ({\rm for~large~ }\Delta)~.
\ee
In this expression $\gamma$ is the opportune value in the second row
of Table I connecting $T_B$ to the numerical estimate of $T_{rh}$,
while $\beta\simeq 2.5$ is the corrective factor at large $\Delta$
determined in Section \ref{comparison}. Subsequently, the horizontal
position of the $\tilde\Delta$ curve is given by
\be
\label{tauBound}
 \tau \lesssim 7.7\times 10^{17}\GeV\frac{\gamma^2}{T_B^2\sqrt{g_{rh}}}
\qquad ({\rm for~large~ }\Delta)~.
\ee

Afterwards, one needs to fix the vertical position of the
$\tilde\Delta$ curve. This is furnished by Fig.~2 because for
$\tau\rightarrow\infty$ we have
\be 
\label{DeltaTau}
\tilde\Delta\approx\Delta\leq \eta^{max}/\eta^{exp}\qquad ({\rm
  for~large~ }\tau)~.
\ee

In conclusion, the compatibility of late-time entropy injection with
successful baryogenesis can be calculated by using simple
arithmetic. Some concrete examples will clarify how this procedure can
be applied in realistic scenarios.

\section{Explicit Applications}
\label{BarBounds}
The procedure we have just presented is model independent except but
the assumptions:
\begin{enumerate}[{  ~~\it i)}]
\item the $X$ decay follows Eq.~(\ref{bltz}), does not induce a $B$
  asymmetry and its products thermalize fast.
\end{enumerate}
Moreover it is opportune to have:
\begin{enumerate}[{\it ii)}]
\item the BAU production is much faster than the entropy
  injection.
\end{enumerate}

Therefore, the described procedure allows for investigating the
compatibility between any specific baryogenesis mechanism and any
particular $X$ field satisfying the conditions {\it i} and (possibly)
{\it ii}.  As explicit applications, we will analyze some concrete
mechanisms embedded in the MSSM whose $g(T)$ is calculated taking the
scalar [fermion] content at $\mathcal O(1)$ [$\mathcal O(0.1)$] TeV
for definiteness.

\subsection{Some baryogenesis mechanisms}
Here we apply our results to electroweak baryogenesis and thermal
(resonant and non resonant) leptogenesis in presence of a generic $X$
field. For each baryogenesis framework we first review the estimate of
$\eta^{max}$~\footnote{We take the estimates of the literature
  assuming standard cosmology but in principle the evaluation of
  $\eta^{max}$ could change when the BAU is produced in the
  $X$-dominated era. However, for our concrete baryogenesis examples,
  Refs.~\cite{Chun:2007np, Prok} find small modifications of $\eta^{max}$
  and $T_B$ so that we can use the standard results.}  and then we
calculate the parameter region where the $X$ field does not destroy
the BAU. The results will be summarized in Fig.~5, in which also the
BBN bound and some $\Delta$ contour (dotted dashed) curves for the
MSSM are reported.


\subsubsection{Electroweak Baryogenesis}
\label{subEWBG}
In electroweak baryogenesis the observed BAU is produced during the electroweak
symmetry breaking \cite{cline}. In this scenario the departure from
thermal equilibrium is achieved by a first order electroweak phase
transition (EWPT). In such a case the transition proceeds via
nucleation of bubbles containing the electroweak broken phase and the movement
of the bubble breaks locally the thermal equilibrium conditions. Then,
just in front of the expanding bubbles, $C$ and $CP$ violating
interactions generate a left-handed asymmetry that $SU(2)_L$
sphalerons transform to a $B$ asymmetry entering the bubble. However,
the formation of the $B$ asymmetry is not enough: it is also necessary
to preserve it till today. This occurs if the first order EWPT is {\it
  strong}~\footnote{The EWPT is strong when $SU(2)_L$ sphalerons are
  out-of-equilibrium inside the bubbles. Unless of subtle
  circumstances \cite{Prok, magnet} not considered here, this happens
  for $v(T_B)/T_B \gtrsim 0.7$ where $v\equiv v(T=0)=174\,\GeV$ and
  $v(T_B)$ is the vacuum expectation value of the (SM-like) Higgs at
  the temperature $T_B$ when the phase transition begins
  \cite{cline}.}.

In the SM the EWPT is not strong and the $CP$ violating sources are
too small to form enough $B$ asymmetry \cite{cline}. These problems
are overcome in some extensions of the SM. Some non-supersymmetric
scenarios have been considered~\cite{Espinosa:2007qk} but at present
the analyses of them are not so developed to provide precise estimates
of the maximal $B$ asymmetry these models can produce. Instead
accurate predictions exist for supersymmetric extensions.

The MSSM can reproduce the observed BAU if the gaugino-Higgsino sector
is at the electroweak scale~\cite{CQW,Cirigli, Carena:2000id,Lee:2004we,
  Konstandin:2005cd}, the SM-like Higgs and right-handed stop are
light ($m_h\lesssim 127$~GeV, $m_{\tilde t_R}\lesssim 120$~GeV) and
the left-handed stop is heavy ($m_{\tilde t_L}\gtrsim 6.5$~TeV)
\cite{Carena:2008rt}. Moreover, the $B$ asymmetry formation typically
starts at $T_B\simeq125$ GeV and ends after a few
GeV~\cite{Carena:2008rt}. Finally, concerning $\eta^{max}$, different
treatments of $CP$-violating sources and flavor effects
exist~\cite{Carena:2000id,Lee:2004we, Konstandin:2005cd} and they lead
to $B$ asymmetries that may differ by almost one order of
magnitude. Choosing the intermediate result we can consider
$\eta^{max}\simeq10\, \eta^{exp}$ \cite{Carena:2000id, Carena:2008rt}.

\begin{figure}
\vspace{-5mm}
\begin{center}
\includegraphics[width=0.47\textwidth]{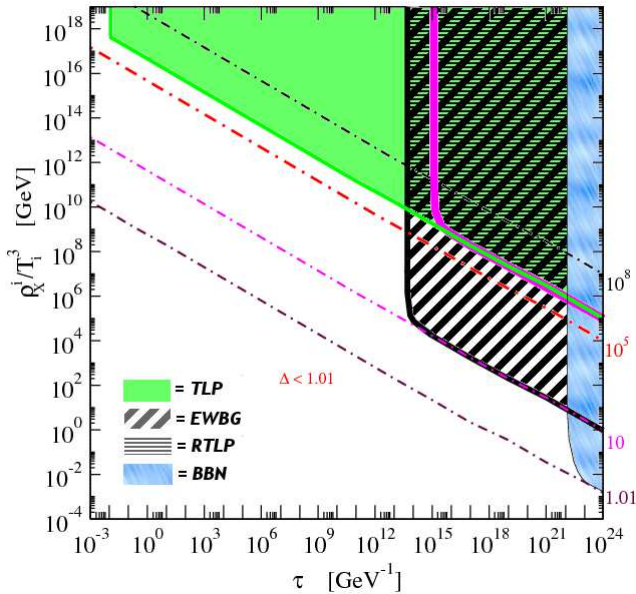}
\caption{Numerical contours (dashed dotted lines) of $\Delta =1.01,
  10, 10^5, 10^8$ (labels on the right) as function of
  $\rho^i_X/T_i^3$ and $\tau$ for MSSM degrees of freedom. Regions
  excluded by BBN, electroweak baryogenesis (EWBG), thermal
  leptogenesis (TLP) and resonant thermal leptogenesis (RTLP) are
  filled as reported in the legend.}
\label{figure-5}
\end{center}
\end{figure}

In conclusion, the parameter space where electroweak baryogenesis in
the MSSM is compatible with the $X$ decay is given by the constraint
\be
\label{EWBG}
\tilde \Delta(T_{B}\simeq 125 {\GeV})\lesssim10~,
\ee
in which the condition {\it ii} is fulfilled because the $B$ asymmetry
is produced in a narrow temperature interval ${\mathcal O}(1 \GeV)$.

The numerical solution of Eq.~(\ref{EWBG}) gives the area labelled
EWBG in Fig.~5. Its border curve could be calculated by the analytic
procedure expressed in Section \ref{III}. In fact, the position of the
vertical part of the curve is given by Eq.~(\ref{tauBound}) yielding
$\tau\lesssim 5\timess10^{13} \GeV^{-1}$ for
$g_{rh}=g(T_{rh}=92\GeV)\simeq 100$ and $\gamma=3.4$, while at much
larger $\tau$ the line must coincide with the bound $\Delta<10$
because of Eq.~(\ref{DeltaTau}). Instead, in the intermediate regime
of $\tau$ the curve is deduced by the universal shape of the
$\tilde\Delta$ lines of Fig.~4.

Before concluding, some words about electroweak baryogenesis in
non-minimal supersymmetric extensions are in order. The estimate
$\eta^{max}\hspace{-1mm}\approx\hspace{-1mm}10\eta^{exp}$ holds
roughly correct also in extensions of the MSSM where the extra content
couples weakly to the Higgs sector and no new large sources of $CP$
violation are introduced. This is not the case if an extra singlet is
added to the MSSM as it is performed in Ref.~\cite{Huber:2006ma}. In
such a extension $\eta^{max}$ can be enhanced approximatively by a
factor 5. Moreover, other modifications of the minimal setup can
further enhance $\eta^{max}$, as for instance in the extension called
Beyond-the-MSSM where $\eta^{max}$ may increase by a further order of
magnitude \cite{BMSSM}. Here we do not explicitly repeat the analysis
for all these variations but their compatibility with a late-time
entropy injection could be easily figured out algebraically as we have
just performed for the MSSM.


\subsubsection{Thermal Leptogenesis} 
Leptogenesis is one of the attractive scenarios to explain the BAU
\cite{yanagida, lep_review}. Its central idea is that in a
$B$-symmetric but Lepton($L$)-asymmetric Universe it is possible to
produce the BAU thanks to electroweak sphalerons that equilibrate the
$B$+$L$ asymmetry without changing $B$--$L$. In thermal leptogenesis
the required initial $L$ asymmetry can be obtained for instance in the
Type-I seesaw framework~\cite{seesaw} where one singlet right-handed
neutrino $N$ per family is added to the SM~\footnote{Our conclusions
  do not change for Type-II and Type-III~\cite{otherTypes}.}. These
fields have the interactions
\begin{equation}
\label{Linteract}
\mathcal{L} \supset \frac{1}{2} (M_N)_{ii} N_i N_i + y_{ij} N_i \bar{\ell}_j i \tau_2 H^* + {\rm h.c.}\;,
\end{equation}
in which $l$ ($H$) is the SM lepton (Higgs) and indeces run over
families. They not only give rise to a net $L$ asymmetry in the early
Universe by out-of-equilibrium decays but also generate sub-eV
neutrino masses via the canonical seesaw mechanism as required by the
neutrino oscillation data \cite{GonzalezGarcia:2010er}. This mass
turns out to be $m_\nu \approx |y|^2 v^2/M_N$ yielding $M_N\sim
10^{14}$ GeV for $|y|\sim 1$ and $m_\nu\sim0.1 \eV$.

Assuming a normal hierarchy in the heavy neutrino sector, the $CP$
asymmetry of the decay of the lightest right-handed neutrino $N_1$ is
given by \cite{type-I-bound}
\begin{equation}
\label{CPtype-I}
  |\epsilon^{I}| = \frac{3 M_{N_1}}{16 \pi v^2}
  \sqrt{\Delta m^2_{\rm atm}} ~\sin \delta ~,
\end{equation}
where $\Delta m^2_{\rm atm}$ is the atmospheric mass scale of light
neutrinos \cite{GonzalezGarcia:2010er} and $\delta$ is the effective
$CP$ violating phase. 

The $L$ asymmetry, which is mostly
produced by the decay of $N_1$, is given by $Y_L = \epsilon^I Y_{N_1}
W$ where the thermal wash out $W$ takes into account the effect of
interactions reducing the created $L$ asymmetry, as for instance $\bar
l H \leftrightarrow l H^*$. Afterwards, electroweak sphalerons tend to
equilibrate the $B$+$L$ asymmetry and so they convert around half part
of $Y_L$ into $\eta$.  Therefore, in order to determine the maximal
$B$ asymmetry $\eta^{max}$, one sets $\sin\delta=1$ and obtains
\cite{type-I-bound}
\be 
\nonumber
\frac{\eta^{max}}{\eta^{exp}}\approx 4 \times 10^4 \left( \frac{M_{N_1}}{10^{14} \rm GeV}\right)
\frac{\sqrt{\Delta m^2_{\rm atm}}}{0.05 \rm eV} \left( \frac{ 
Y_{N_1} W}{7 \times 10^{-4}} \right)~.
\label{type-I}
\ee
Since commonly the framework is supposed to be embedded in a grand
unified theory and moreover the Universe typically never reached
temperatures high enough to thermally generate neutrinos beyond the
GUT scale, we can consider $ 10^{16} \gtrsim
M_{N_1}\hspace{-1mm}\GeV^{-1} \gtrsim 10^{9}$, where the lower bound
comes from requiring enough BAU in Eq.~(\ref{type-I}). 

Now we can analyze the $B$-asymmetry dilution due to $X$
decay~\footnote{An analysis focused on solving the gravitino problem
  in a particular framework of thermal leptogenesis and late-time
  decay is presented in Ref.~\cite{Hasenkamp:2010if}.}. Taking
$T_B\approx M_{N_1}$, the case $M_{N_1}\simeq 10^{9} \GeV$ gives
$\eta^{max}/\eta^{exp}=\mathcal O(1)$ and yields then
\be
\label{TLP-1}
\tilde \Delta(T_B\simeq10^9 \GeV) \lesssim {\cal O}(1)~.
\ee
Eq.~(\ref{TLP-1}) permits any $\Delta$ for $\tau\lesssim
10^{-1}\GeV^{-1}$ but does not provide a sharp bound on $\tau$ because
the condition {\it ii} is not guaranteed.  Similarly for
$M_{N_1}\simeq 10^{16} \GeV$, which yields
$\eta^{max}/\eta^{exp}=\mathcal O (10^6)$, we have the constraint
\be
\label{TLP-2} 
\tilde \Delta(T_B\simeq10^{16} \GeV)\lesssim {\cal O}(10^6)~,
\ee
that allows any dilution $\Delta\lesssim 10^6$ independently of
$\tau$.

The union of the regions fulfilling (\ref{TLP-1}) and (\ref{TLP-2})
gives the parameter space where thermal leptogenesis is compatible
with the $X$ decay. The numerical result is shown in Fig.~5 (green
area labelled TLP) and could be reproduced algebraically by the
procedure explained in Section \ref{III} and already applied for
electroweak baryogenesis.

\subsubsection{Resonant thermal leptogenesis}
The $CP$ asymmetry in Eq.~(\ref{CPtype-I}) requires very heavy
additional neutrinos to produce the BAU. For this reason thermal
leptogenesis with normal neutrino hierarchy provides no experimental
evidence at current achievable energies. This unappealing feature is
avoided in resonant thermal leptogenesis where extra neutrinos at the
electroweak scale may yield BAU and detectable signatures at the same
breath \cite{Franceschini:2008pz}.

The key idea of resonant
leptogenesis~\cite{resonant_lep1,resonant_lep2} is that the
self-energy effects dominate the leptonic asymmetries when the mass
splitting between the right-handed neutrinos $N$ is much less than
their masses. If the mass splitting between these fields is comparable
to their decay widths, the $CP$ asymmetry gets enhanced resonantly.

Focusing on type-I leptogenesis (our conclusions would not change for
type II and III), let us consider two singlet Majorana fields $N_1$
and $N_2$ of masses $M_1$ and $M_2$. Their Yukawa interaction $h_{ij}
N_i \bar{\ell}_j H$ allows for decays to SM lepton $\ell$ and Higgs
doublet H whose $CP$ asymmetry is given by~\cite{plumacher}
\begin{equation}
\epsilon_1 = \frac{{\rm Im} (h^\dagger h)_{12}^2}{8\pi (h^\dagger h)_{11}}
\frac{(M_1^2-M_2^2)M_1 M_2}{(M_1^2-M_2^2)^2+ (M_2 \Gamma_2 - M_1\Gamma_1)^2}~.
\end{equation}
Hence, assuming $M_1\sim M_2$ and $M_1-M_2 \sim \Gamma_1-\Gamma_2$ we
can achieve $\epsilon_1 \sim {\cal O}(1)$ so that a large $L$
asymmetry $Y_L$ can be produced even if the initial $N_1$ and $N_2$
abundance $Y_N$ is small. Finally, $SU(2)_L$ sphalerons can convert
$Y_L$ to $Y_B$ if the lepton asymmetry is produced enough before
$T\simeq130\,$GeV when sphalerons decouple
\cite{Burnier:2005hp}. Typically, this happens if $M_1\gtrsim 250
\GeV$~\cite{resonant_lep2}. In such a case it turns out $Y_B\simeq 0.5
\times \epsilon_1 Y_N W$, where it is reasonable to consider $Y_N W
=\mathcal O(10^{-4})$ \cite{Franceschini:2008pz} yielding
$\eta^{max}={\cal O}(10^{-4})$ for $\epsilon_1 \sim {\cal O}(1)$.

In conclusion, in resonant leptogenesis it turns out to be
$Y_B^{max}\approx \mathcal O (10^6) \eta^{exp}$. Moreover, in order to
favor the compatibility with the late-time decay, $T_B$ has to be as
low as possible. Since it is $T_B\approx M_1$, the most favorable
choice is $M_1\approx 250\,$GeV. Subsequently, resonant leptogenesis
is compatible with the $X$ decay when the constraint
\be
\label{RTLP}
\tilde\Delta(T_{B}\simeq 250 \GeV)\lesssim \mathcal O (10^6)~, 
\ee
is fulfilled. Observe that this constraint can be considered as a
sharp bound since the condition {\it ii} is satisfied due to the short
temperature interval during which the $B$ asymmetry is generated
($250\!\gtrsim\!  T\!/\!\!\GeV \!\gtrsim\!  130$).

By solving Eq.~(\ref{RTLP}) numerically, one finds that the excluded
region is the area labelled RLPT shown in Fig.~5.  As already checked
for the previous baryogenesis mechanisms, the excluded region could be
easily determined algebraically by the procedure of Section
\ref{III}. Indeed its vertical border is due to the bound $\tau\gtrsim
1\times10^{15} \GeV^{-1}$ coming from Eq.~(\ref{tauBound}) with
$g_{rh}=g(T_{rh}=18\GeV)\simeq 90$ and $\gamma=33$, its lower part
corresponds to the constraint $\tilde\Delta\simeq\Delta<10^6$ shown in
Fig.~2, and the curved part is deduced by copying the shape of the
$\tilde\Delta$ lines of Fig.~4.


\subsection{ $X$ field as a modulus}
In the examples we have just considered the field producing the
late-time entropy is not specified. Now we analyze the particular case
of the $X$ field being a modulus. Let us assume it to be
gravitationally coupled to the SM thermal bath via a dimension-five
operator so that its lifetime is given by
\be
\label{tauosc}
\tau =  \frac{2\pi}{\alpha_X} \left( \frac{m_P^2}{M_X^3}\right)~,
\label{decay_rate}
\ee 
where $\alpha_X$ spans from ${\cal O}(1)$ to ${\cal
  O}(10^{-2})$ depending on the non-renormalizable
coupling~\cite{dutta&sinha}. 

In an expanding Universe, when the Hubble scale is roughly equal to
the modulus mass, $H\approx M_X $, the modulus enters into the
oscillating regime.  By taking the initial amplitude of oscillations
of $X$ to be ${\mathcal X}_i$, the energy density $\rho_X^i$ can be
expressed as
\be
\label{mod_rXi}
\rho_X^i = M_X^2 \XX^2~.
\ee
Moreover, when the field starts oscillating, the Universe is radiation
dominated and its temperature $T_i$ is
\be
\label{mod_Ri}
T_i^4= \frac{90  ~m_P^2 M_X^2}{\pi^2 g_i} ~.
\ee
%
%
\begin{figure}
\begin{center}
\vspace{-5mm}
\includegraphics[width=0.5\textwidth]{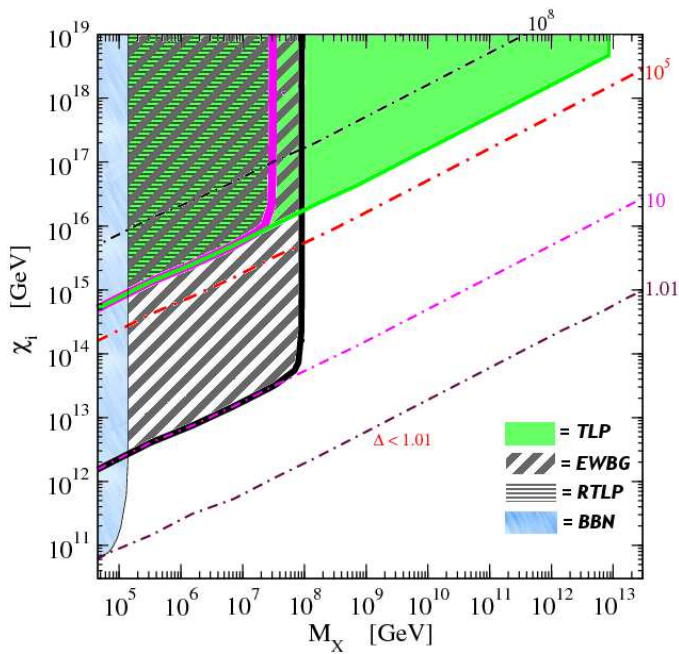}
\caption{Implications of the analysis summarized in Fig.~5 for decay
  of muduli with coupling $\alpha_X=1$. Numerical contours (dotted
  dashed curves) of $\Delta =1.01, 10, 10^5, 10^8$ (labels on the
  right and top) as function of $\XX$ and $M_X$.  Regions excluded by
  BBN, electroweak baryogenesis (EWBG), thermal leptogenesis (TLP) and
  resonant thermal leptogenesis (RTLP) are filled as reported in the
  legend.}
\label{figure-6}
\end{center}
\end{figure}
%
Subsequently, we can easily convert the model independent bounds of
the previous sections to the case of $X$ as a modulus. For instance,
the general bounds on $\tau$ and $\rho_X^i/T_i^3$ presented in Fig.~5
can be re-expressed as constraints on $\XX$ and $M_X$, as shown in
Fig.~6. 

Using Eqs.~(\ref{tauosc})-(\ref{mod_Ri}), the sudden decay
approximations (\ref{Te}), (\ref{Delta}) and (\ref{Trh}) become
\bea
T_e^4&=&\frac{10}{9 \pi^2 g_i} \frac{\XX^8 M_X^{2}}{m_P^{6}}~,
\\
\label{TrhMod}
T_{rh}^4&=&\frac{10}{\pi^4 g_{rh}} \frac{\alpha_X^2 M_X^{6}}{m_P^2} ~,
\\
\Delta&=&\sqrt[4]{\frac{\pi^2~g_{rh}}{9 ~ g_i}}
\frac{\XX^{2}}{\sqrt{\alpha_X} m_P M_X} ~,
\eea
and the BBN bound $T_{rh}\!\geq\!4\MeV$ is converted to
\be
 M_X \gtrsim 7 \times 10^4 ~ (1/\alpha_X)^{1/3} {\rm GeV} \qquad
({\rm analytic}) ~, 
\ee
which is slightly weaker than the numerical one. However, the BBN
bound is not the strongest constraint in the regime of large
oscillations. In such a regime, to avoid the BAU wash out, we need
$M_{\!X}\!\sqrt[3]{\alpha_X}> {\cal O}(10^{13})$~GeV for thermal
leptogenesis, $M_{\!X}\!\sqrt[3]{\alpha_X}>{\cal O}(10^{8})$~GeV for
electroweak baryogenesis and $M_{\!X}\!\sqrt[3]{\alpha_X}>{\cal
  O}(10^{7})$~GeV for resonant leptogenesis.

\section{Conclusions}
%
There exist many theories beyond the standard model of particle
physics that predict cosmologically long-lived fields. Even though the
initial energy density of these fields may be many orders of magnitude
less than the one of radiation, at later times they can dominate the
expansion of the Universe and dump sizeable amount of entropy during
their decay. Subsequently these fields dilute the pre-processed $B$
asymmetry and may alter the primordial element abundances.

In order to constrain the above theories, in this paper we analyzed
the entropy and the subsequent re-reheating that a generic long-lived
field $X$ produces. We solved the decay equations numerically and we
checked their consistency with the analytic sudden-decay
approximation. The result can be parameterized by lifetime and initial
energy density of the $X$ field. We found that for small entropy
dilution $\Delta$ the analytic approximation badly fails because in
this regime there is no clear $X$-dominated epoch. Instead for $\Delta
\gtrsim 3$ the analytic and numerical calculations disagree by less
than 10\% in the entropy estimates but by a factor $\sim$2.5 in the
evaluation of $T_{rh}$, the temperature at the end of the
dilution. However, the errors of the analytic approach can be overcome
by applying some corrective factors which, among other effects,
strength slightly the BBN upper bound on the $X$ lifetime.

The revised control on entropy dilution and re-reheating temperature
allowed us for investigating the effect of the $X$ decay on the $B$
asymmetry. The analysis was carried out by assuming the mechanism
producing the entropy injection to be independent of the one
generating the BAU. The outcome is a set of conditions on the lifetime
and initial energy density of $X$ as function of when and in what
amount the BAU is produced. These conditions are quite generic so that
they can be easily applied to determine the compatibility of
baryogenesis and late-time decay described by a wide class of
scenarios.

As illustrative examples, we applied our results to some concrete
models.  It turned out that successful MSSM electroweak baryogenesis,
thermal leptogenesis and resonant leptogenesis put strong bounds on
the initial abundances of $X$ fields with lifetime $\tau\gtrsim
5\times10^{13}, 10^{-2}, 10^{15}\,\GeV^{-1}$, respectively. Instead,
for smaller values of $\tau$ the abundance is not constrained (see
Fig.~5). Furthermore, if we take $X$ being modulus as explicit example
of decaying field, the above constraints become bounds on the initial
oscillation amplitude $\XX$, mass $M_X$ and gravitationally-mediated
coupling $\alpha_X$ to the visible sector. In particular, very large
$\XX$ implies excessive dilution of the $B$ asymmetry produced by MSSM
electroweak baryogenesis, thermal leptogenesis and thermal resonant
leptogenesis for $M_{\!X}\!\sqrt[3]{\alpha_X} \gtrsim 10^8, 10^{13},
10^{7}\,\GeV$, respectively (see Fig.~6).

\begin{acknowledgments}
  We are grateful to A.~Mazumdar for inspiring discussions and
  suggestions in several stages of the paper. We also thank T.~Hambye
  and C.~Ringeval for comments on the manuscript and on CMB
  bounds. G.N.~thanks M.~Carena, T.~Konstandin, M.~Quir\'os and
  C.~Wagner for many discussions on electroweak baryogenesis.  The
  work of G.N.~and N.S.~is supported by IISN and the Belgian Science
  Policy (IAP VI-11).
\end{acknowledgments}

\appendix
\section*{Appendix:~~ Relaxing the BBN bound}
\label{appA}

Apart from some special exceptions~\cite{Jedamzik:2009uy}, the
observed primordial element abundances cannot be explained when
sizeable entropy injections occur during the BBN epoch. On the
contrary, the success of the standard BBN model is not spoiled
for~\cite{Hannestad:2004px}
\be
\label{tooConser2}
T_{rh}\geq T_{BBN}\equiv 4\,\MeV~, \ee
where the $X$-decay hadronic channels are supposed suppressed.

In our numerical analysis we implement the reheating temperature as
\be
\label{Trh_defA}
\frac{S(T\ll T_{rh})}{S(T_{rh})}\equiv 1+\xi~,
\ee
with $\xi=0.01$.  Subsequently $T_{rh}$ is the temperature of the
thermal bath when the 99\% of the entropy due to the $X$ decay has
been injected.

Applying the definition (\ref{Trh_defA}) in the constraint
(\ref{tooConser2}) is consistent with BBN. Indeed, even in the
pathological combination of baryogenesis and entropy mechanisms
leading to $T_{rh}= T_{BBN}$ and $\eta(T_{BBN})\simeq 8.4\times
10^{-11}$ [the minimal value allowed by Eqs.~(\ref{BBN}) and
(\ref{WMAP})], the remaining 1\% of entropy injection after $T_{BBN}$
dilutes $\eta$ by an amount that is practically negligible for the
experimental constraints.

However, the convention $\xi=0.01$ is often too conservative. For
instance, in the case with $T_{rh}= T_{BBN}$ and $\eta(T_{BBN})\simeq
9.2\times 10^{-11}$, the maximal allowed entropy dilution after
$T_{BBN}$ is the ratio between the upper and lower bounds of
$\eta^{exp}$. In such a case, by taking the allowed extrema of
(\ref{BBN}) and (\ref{WMAP}), the BBN bound can be implemented by
(\ref{tooConser2}) where $T_{rh}$ is redefined through
(\ref{Trh_defA}) with $\xi=0.10$\,. On the other hand, if possible
systematic errors due to priors were taken into account in the WMAP
analysis, the experimental bound (\ref{WMAP}) would be comparable with
(\ref{BBN})~\cite{pdg} and the BBN bound could be further relaxed by
redefining $T_{rh}$ by (\ref{Trh_defA}) with $\xi=0.27$.

As a consequence, these different choices of $\xi$ slightly relax the
BBN bound shown in the figures of the paper.  For instance, the
convention $\xi=0.27$ weakens the BBN constraint on $\rho_X^i/T_i^3$
(on $\tau$) by a factor $\sim$10 $(\sim$3). In any case for the
possible values $0.01<\xi<0.27$ the numerical analysis still provides
a BBN bound that is stronger than the one
obtained analytically by Eq.~(\ref{Trh}).

\end{document}